\documentclass[11pt]{article}
\usepackage{amsmath,amssymb,amstext,bm,amsthm}
\usepackage{amsfonts}
\usepackage{mathrsfs}
\usepackage{color}
\usepackage{pstricks,pst-plot,pstricks-add,pst-math}
\newtheorem{lemma}{Lemma}
\newtheorem{theorem}{Theorem}
\newtheorem{definition}{Definition}
\newtheorem{proposition}{Proposition}
\theoremstyle{remark}
\newtheorem{remark}{Remark}
\newtheorem{example}{Example}

\newcommand{\eps}{\epsilon}

\newcommand{\R}{\mathbb R}

\newcommand{\X}{\underline{X}}
\newcommand{\G}{\widehat{G}}
\newcommand{\tbeta}{\widetilde{\beta}}
\renewcommand{\phi}{\varphi}

\title{ {Reconstruction of Markov Random Fields from Samples:\\
         Some Observations and Algorithms}
\author{ Guy Bresler\thanks{Department of Electrical Engineering and Computer Sciences, U.C. Berkeley. Email: {\tt gbresler@eecs.berkeley.edu}. Supported by a Vodafone US-Foundation fellowship.} \and
Elchanan Mossel\thanks{Dept. of Statistics and Dept. of Electrical Engineering and Computer Sciences, U.C. Berkeley. E-mail: {\tt
        mossel@stat.berkeley.edu}. Supported by a Sloan fellowship
  in Mathematics, by NSF Career award DMS-0548249, NSF grant
  DMS-0528488 and ONR grant N0014-07-1-05-06}
\and Allan Sly\thanks{Dept. of
Statistics, U.C. Berkeley. Email: {\tt sly@stat.berkeley.edu  } Supported by NSF grants DMS-0528488 and
DMS-0548249 }
        }}

\begin{document}
\maketitle
\begin{abstract}
Markov random fields are used to model high dimensional
distributions in a number of applied areas. Much recent interest has
been devoted to the reconstruction of the dependency structure from
independent samples from the Markov random fields. We analyze a
simple algorithm for reconstructing the underlying graph defining a
Markov random field on $n$ nodes and maximum degree $d$ given
observations. We show that under mild non-degeneracy conditions it
reconstructs the generating graph with high probability using
$\Theta(d \epsilon^{-2}\delta^{-4} \log n)$ samples where $\epsilon,\delta$ depend on the local interactions. For most local interaction $\eps,\delta$ are of order $\exp(-O(d))$.

Our results are optimal as a function of $n$ up to a multiplicative
constant depending on $d$ and the strength of the local interactions.
Our results seem to be the first results for general models that guarantee
that {\em the} generating model is reconstructed. Furthermore, we
provide explicit $O(n^{d+2} \epsilon^{-2}\delta^{-4}  \log n)$ running time bound. In cases
where the measure on the graph has correlation decay, the running time
is $O(n^2 \log n)$ for all fixed $d$. We also discuss the effect of
observing noisy samples and show that as long as the noise level is
low, our algorithm is effective. On the other hand, we construct an
example where large noise implies non-identifiability even for
generic noise and interactions. Finally, we briefly show that in
some simple cases, models with hidden nodes can also be recovered.

\end{abstract}

\section{Introduction}
In this paper we consider the problem of reconstructing the graph
structure of a Markov random field from independent and
identically distributed samples. Markov random fields (MRF) provide
a very general framework for defining high dimensional distributions
and the reconstruction of the MRF from observations has attracted
much recent interest, in particular in biology, see
e.g.~\cite{Friedman} and a list of related references~\cite{list}.

\subsection{Our Results}
We give sharp, up to a multiplicative constant, estimates for the
number of independent samples needed to infer the underlying  graph
of a Markov random field of bounded degree. In Theorem~\ref{t:lowerBound}
we use a simple information-theoretic
argument to show that $\Omega(d\log n)$ samples are required to
reconstruct a randomly selected graph on $n$ vertices with maximum
degree at most $d$. Then in Theorems~\ref{t:reconstructionNew}
and~\ref{t:reconstruction} we propose two algorithms for reconstruction
that use only $O(d\epsilon^{-2}\delta^{-4}\log n)$ where $\epsilon$ and $\delta$ are lower bounds on marginal distributions in the neighbourhood of a vertex. Under mild non-degeneracy conditions $\epsilon,\delta=\exp(-O(d))$
and for some models $\epsilon,\delta=\hbox{poly}^{-1}d$. Examples of the later model include the hardcore model with fugacity $\lambda=\Theta(\frac1d)$.
 Our main focus is on the reconstruction of sparse MRFs case where $d$ is fixed in which case $\epsilon$ and $\delta$ are constant.
The two theorems differ in their
running time and the required non-degeneracy conditions.
It is clear that non-degeneracy conditions
are needed to insure that there is a unique graph associated with
the observed probability distribution.

In addition to the fully-observed setting in which samples of all
variables are available, we extend our algorithm in several
directions. In Section~\ref{sec:noisyetc}
we consider the problem of noisy observations.
In subsection~\ref{subsec:nonid} we
show by way of an example that if some of the random variables are
perturbed by noise then it is in general impossible to reconstruct
the graph structure with probability approaching 1. Conversely, when
the noise is relatively weak as compared to the coupling strengths
between random variables, we show that the algorithms used in
Theorems~\ref{t:reconstructionNew}
and~\ref{t:reconstruction}
reconstruct the graph with high probability.
Furthermore, we study the problem of
reconstruction with partial observations, i.e. samples from only a
subset of the nodes are available. In Theorem~\ref{t:missing_vertex}
we provide sufficient conditions on
the probability distribution for correct reconstruction.

Chickering \cite{C95} showed that maximum-likelihood estimation of the underlying graph of a Markov random field is NP-complete. This does not contradict our results which assume that the data is generated from a model
(or a model with a small amount of noise). Although the algorithm we propose runs in time polynomial in the
size of the graph, the dependence on degree (the run-time is $O(
n^{d+2} \epsilon^{-2}\delta^{-4} \log n)$) may impose too high a computational cost for some
applications.   Indeed, for some Markov random fields exhibiting a decay of correlation
a vast improvement can be realized: a
modified version of the algorithm runs in time $O(dn^2 \epsilon^{-2}\delta^{-4} \log n)$.
This is proven in Theorem~\ref{t:reconstructionDecay}.


\subsection{Related Work}
Chow and Liu \cite{CL68} considered the problem of estimating Markov random fields whose underlying graphs are trees, and provided an efficient (polynomial-time) algorithm based on the fact that in the tree case maximum-likelihood estimation amounts to the computation of a maximum-weight spanning tree with edge weights equal to pairwise empirical mutual information. Unfortunately, their approach does not generalize to the estimation of Markov random fields whose graphs have cycles. Much work in mathematical biology is devoted to reconstructing tree Markov fields when there
are hidden models. For trees, given data that is generated from the model, the tree can be reconstructed efficiently
from samples at a subset of the nodes given mild non-degeneracy conditions.
See~\cite{ErStSzWa:99a,Mossel:07,DaMoRo:06} for some of the most recent and tightest results in this setup.

The most closely related works are \cite{Koller} and \cite{Wainwright}.  These can be compared in terms of sampling complexity, running time as well as the generality of the models to which they apply.  These are summarized in the Table below.
The first line refers to the type of models that the method cover: Does the model allow clique interactions of just edge interactions? The next two lines refer to requirements on the strength of interactions: 
are they not required to be too weak / are only edges with strong interactions returned? are they not required to be too strong? The next line refers to the hardness of verifying if a given model satisfies the conditions of the algorithm (where X denoted that the verification is exponential in the size of the model). The following line refers to the following question: is there a guarantee that the generating model is returned with high probability. The final two lines refers to computational and sampling complexity where $c_d$ denotes constants that depend on $d$.

\medskip
\noindent
\begin{tabular} {|l|c|c|c|c|}\hline
{\bf Method} & {\bf AKN~\cite{Koller}} & {\bf WRL~\cite{Wainwright}} & {\bf Alg} & {\bf High Temp Alg}\\ \hline
\hline Cliques           & $\surd$ & X & $\surd$ & $\surd$ \\
\hline No Int. Low. Bd.  & $\surd$ & X & X & X \\
\hline No Int. Upp. Bd.  & $\surd$ & X & $\surd$ & X \\
\hline Verifiable Conds. & $\surd$ & X & $\surd$ & $\surd$ \\
\hline Output Gen. Model & X & $\surd$ & $\surd$ & $\surd$ \\
\hline Comp. Compl.      & $n^{O(d)}$   & $n^5$ & $n^{O(d)}$ & $c_d n^2 \log n$\\
\hline Sampl. Compl.     & $n^{O(d)}$ & $poly(d) \log n$ & $c_d \log n$ & $c_d \log n$\\
\hline \hline
\end{tabular}

\medskip
Abbeel, \emph{et al} \cite{Koller} considered the problem of
reconstructing graphical models based on factor graphs, and proposed
a polynomial time and sample complexity algorithm. However, the goal
of their algorithm was not to reconstruct the true structure, but
rather to produce a model whose distribution is close in Kullback-Leibler
divergence to the true distribution. In applications it is often of
interest to reconstruct the true structure which give some insights
into the underlying structure of the inferred model.

Note furthermore that two networks that differ only in
the neighborhood of one node will have $O(1)$ KL distance. Therefore, even in
cases where it is promised that the KL distance between
the generating distribution and any other distribution defined by another graph
is as large as possible,
the lower bounds on the KL distance is $\Omega(1)$. Plugging this into the bounds in~\cite{Koller} yields a polynomial sampling complexity in the size of the network in order to find the
generating network compared to our
logarithmic sampling complexity. For other work
based on minimizing the KL divergence see the references in
\cite{Koller}.

The same problem as in the present work (but restricted to the Ising model) was
studied by Wainwright, \emph{et al} \cite{Wainwright}, where an
algorithm based on $\ell_1$-regularization was introduced. The algorithm presented is efficient also for dense graphs with running time
$O(n^5)$ but is applicable only in very restricted settings.
The work only applies to the
Ising model and more importantly only models with edge interactions (no larger cliques are allowed).
The most important restrictions are the two conditions in the paper (A1 and A2).
{Condition A1 requires (among other things) that the ``covariates [spins] do not become overly dependent''. Verifying when the conditions holds seems hard.} However, it is easy to see that this condition fails for standard models
such as the Ising model on the lattice or on random $d$-regular graphs when the model is at low temperatures, i.e. for $\beta > \frac12\log(1+\sqrt{2})$ in the case of the two dimensional Ising model and $\beta > \tanh^{-1}\big(1/(d-1)\big)$ for random $d$-regular graphs.

Subsequent to our work being posted on the Arxiv, Santhanam and Wainwright \cite{SW08}
again considered essentially the problem for the Ising model, producing nearly
matching lower and upper bounds on the asymptotic sampling
complexity. Again their conditions do not apply to the low temperature regime.  Another key difference from our work is that they restrict
attention to the Ising model, i.e. Markov random fields with
pairwise potentials and where each variable takes two values.  Our results are not limited to pairwise interactions and apply to the more general setting of MRFs with potentials on larger cliques.

\section{Preliminaries}
We begin with the definition of Markov random field.
\begin{definition}
On a graph $G=(V,E)$, a \emph{Markov random field} is a distribution $X$ taking
values in $\mathcal{A}^V$, for some finite set $\mathcal{A}$ with
$|\mathcal{A}|=A$, which satisfies the Markov property
\begin{equation}\label{e:MarkovProperty}
P(X(W),X(U)|X(S))=P(X(W)|X(S))P(X(U)|X(S))
\end{equation}
when $W,U$ and $S$ are disjoint subsets of $V$ such that every path
in $G$ from $W$ to $U$ passes through $S$ and where $X(U)$ denotes
the restriction of $X$ from $\mathcal{A}^V$ to $\mathcal{A}^U$ for
$U\subset V$.
\end{definition}
Famously, by the Hammersley-Clifford Theorem, such distributions can be written in a factorized form as
\begin{equation}\label{e:gibbsDist}
P(\sigma)=\frac1{Z}\exp\left[ \sum_{a} \Psi_{a}
(\sigma_a) \right]
\end{equation}
where $Z$ is a normalizing constant, $a$ ranges over the cliques in $G$, and
$\Psi_{a}\colon\mathcal{A}^{|a|}\rightarrow \mathbb{R}\cup \{-\infty\}$ are
functions called \emph{potentials}.

The problem we consider is that of reconstructing the graph $G$, given $k$ independent samples
 $\underline{X}=\{X^1,\ldots,X^k\}$
from the model. Denote by $\mathcal{G}_d$ the set of labeled graphs
with maximum degree at most $d$. We assume that the graph $G\in
\mathcal{G}_d$ is from this class.  A structure estimator (or reconstruction algorithm)
$\widehat{G}:\mathcal{A}^{kn} \to \mathcal{G}_d$ is a map from the
space of possible sample sequences to the set of graphs under
consideration. We are interested in the asymptotic relationship between the number of nodes in the graph, $n$, the maximum degree $d$, and the number of samples $k$ that are required. An algorithm using number of samples $k(n)$ is deemed successful  if in the limit of large $n$ the probability of reconstruction error approaches zero.

\section{Lower Bound on Sample Complexity}
Suppose $G$ is selected uniformly at random from $\mathcal{G}_d$.
The following theorem gives a lower bound of $\Omega(d \log n)$ on the number of samples necessary to reconstruct the graph $G$. The argument is information theoretic, and follows by comparing the number of possible graphs with the amount of information available from the samples.


\begin{theorem}
\label{t:lowerBound}
Let the graph $G$ be drawn according to the uniform
distribution on $\mathcal{G}_d$. Then there exists a constant $c=c(A)>0$ such that if $k\leq c
d\log n$ then for any estimator $\widehat{G}:\X\to \mathcal{G}_d$, the probability of
correct reconstruction is $P(\widehat{G}=G)=o(1)$.
\end{theorem}

\begin{remark}
Note that the theorem above doesn't need to assume anything about
the potentials. The theorem applies for any potentials that are
consistent with the generating graph. In particular, it is valid
both in cases where the graph is ``identifiable'' given many samples
and in cases where it isn't.
\end{remark}

\begin{proof}
To begin, we note that the probability of error is minimized by letting $\G$ be the maximum a posteriori (MAP) decision rule,
\[
\G_{\text{MAP}}(\underline{X}) = \hbox{argmax}_{g\in G}P[G=g|\underline{X}].
\]
By the optimality of the MAP rule, this bounds the probability of error
using any estimator. Now, the MAP estimator $\G_{\text{MAP}}(\underline{X})$ is a deterministic function of
$\underline{X}$.  Clearly, if a graph $g$ is not in the range of
$\widehat{G}$ then the algorithm always makes an error when $G=g$. Let $S$ be the set of graphs in the range of $\G_{\text{MAP}}$, so $P(\text{error}|g\in S^c)=1$. We have
\begin{equation}\begin{split}\label{e:errorLowerBound}
P(\text{error}) &=
    \sum_{g\in \mathcal{G}} P(\text{error}|G=g)P(G=g)
\\&=
    \sum_{g\in S}P(\text{error}|G=g)P(G=g)+\sum_{g\in S^c}P(\text{error}|G=g)P(G=g)
\\&\geq
    \sum_{g\in S^c}P(G=g)
=
    1-\sum_{g\in S} |\mathcal{G}|^{-1}
\\&\geq
    1-\frac{A^{nk}}{|\mathcal{G}|},
\end{split}\end{equation} where the last step follows from the fact that $|S|\leq |\underbar{X}|\leq A^{nk}$.
It remains only to express the number of graphs with max degree at most $d$, $|\mathcal{G}_d|$, in terms of the
parameters $n,d$. The following lemma gives an adequate bound.
\begin{lemma}Suppose $d\leq n^\alpha$ with $\alpha<1$. Then the number of graphs with max degree at most $d$, $|\mathcal{G}_d|$, satisfies \begin{equation} \log |\mathcal{G}_d|=\Omega (nd\log n).\end{equation}
\label{l:graphCounting}
\end{lemma}
\begin{proof} To make the dependence on $n$ explicit, let
$U_{n,d}$ be the number of graphs with $n$ vertices with maximum
degree at most $d$. We first bound $U_{n+2,d}$ in terms of
$U_{n,d,}$. Given a graph $G$ with $n$ vertices and degree at most
$d$, add two vertices $a$ and $b$. Select $d$ distinct neighbors
$v_1,\dots, v_d$ for vertex $a$, with $d$ labeled edges; there are
${n\choose d}d!$ ways to do this. If $v_i$ already has degree $d$ in
$G$, then $v_i$ has at least one neighbor $u$ that is not a neighbor
of $a$, since there are only $d-1$ other neighbors of $a$. Remove
the edge $(v_i,u)$ and place an edge labeled $i$ from vertex $b$ to
$u$. This is done for each vertex $v_1,\dots,v_d$, so $b$ has degree
at most $d$.  The graph $G$ can be reconstructed from the resulting
labeled graph on $n+2$ vertices as follows: remove vertex $a$, and
return the neighbors of $b$ to their correct original neighbors
(this is possible because the edges are labeled).

Removing the labels on the edges from $a$ and $b$ sends at most $d!^2$ edge-labeled graphs of this type
on $n+2$ vertices to the same unlabeled graph. Hence,
the number of graphs with max degree $d$ on $n+2$ vertices is lower
bounded as $$U_{n+2,d}\geq U_{n,d}{n\choose
d}d!\frac{1}{d!^2}=U_{n,d}{n\choose d}\frac{1}{d!}.$$ It follows
that for $n$ even (and greater than $2d+4$)
\begin{equation}U_{n,d}\geq \prod_{i=1}^{n/2}{n-2i\choose d}\frac{1}{d!}\geq
\left({n/2\choose d}\frac{1}{d!}\right)^{n/4}.\label{e:graphCounting1}\end{equation}
 If $n$ is odd, it suffices to note that $U_{n+1,d}\geq U_{n,d}$.
Taking the logarithm of equation \eqref{e:graphCounting1} yields
\begin{equation}
   \log U_{n,d}= \Omega(nd (\log n-\log d))=\Omega (nd\log n),
\end{equation}
assuming that $d\leq n^\alpha$ with $\alpha<1$.
\end{proof}

Together with equation \eqref{e:errorLowerBound}, Lemma
\ref{l:graphCounting} implies that for small enough $c$, if the
number of samples $k\leq c d\log n$, then
\[
P(\text{error})\geq1-\frac{A^{nk}}{|\mathcal{G}|}=1-o(1).
\]
This completes the proof of Theorem \ref{t:lowerBound}.
\end{proof}

\section{Reconstruction}
We now turn to the problem of reconstructing the graph structure of a Markov random field from samples.
For a vertex $v$ we let $N(v)=\{u\in V-\{v\}:(u,v)\in E\}$ denote
the set of neighbors of $v$.  Determining the neighbors of $v$ for
every vertex in the graph is sufficient to determine all the edges
of the graph and hence reconstruct the graph.  We test each
candidate neighborhood of size at most $d$ by using the Markov
property, which states that for each $w\in V-(N(v)\cup \{v\})$
\begin{equation}\label{e:MarkovProperty2}
P(X(v)|X(N(v)),X(w))=P(X(v)|X(N(v)))\, .
\end{equation}

We give two theorems for reconstructing networks; they differ in their
non-degeneracy conditions and their running time. The first one, immediately below, has more stringent non-degeneracy conditions and faster running time.

\subsection{Conditional Two Point Correlation Reconstruction}

\begin{theorem}\label{t:reconstructionNew}
Suppose the graphical model satisfies the following: there exist
$\epsilon,\delta>0$ such that for all $v\in V$, if $U\subset V-\{v\}$
with $|U|\leq d$ and $N(v)\nsubseteq U$ then there exist values
$x_v,x_w,x_w',x_{u_1},\ldots,x_{u_l}$ such that for some $w\in
V-(U\cup\{v\})$
\begin{equation}\begin{split}\label{e:conditionANew}
&\big|P(X(v)=x_v|X(U)=x_{U},X(w)=x_w)
\\&\quad
    -P(X(v)=x_v|X(U)=x_{U},X(w)=x_w')\big| >\epsilon
\end{split}\end{equation}
and
\begin{equation}\begin{split}\label{e:conditionBNew}
&\left|P(X(U)=x_{U},X(w)=x_w)\right|
> \delta,
\\
&\left|P(X(u_1)=x_{U},X(w)=x_w')\right|
> \delta.
\end{split}\end{equation}
Then with the constant $C=\left(\frac{81(d+2)}{\eps^2 \delta^4 2d}+C_1\right)$, when $k>C d\log n$,
there exists an estimator $\hat{G}(\underline{X})$ such that the probability of correct reconstruction is
$P(G=\hat{G}(\underline{X}))=1-O(n^{-C_1})$.
The estimator $\hat{G}$ is efficiently computable in $O(n^{d+2} \log n)$
operations.
\end{theorem}

\begin{remark}Condition \eqref{e:conditionANew}
captures the notion that each edge should have sufficient strength.
Condition \eqref{e:conditionBNew} is required so that we can
accurately calculate the empirical conditional probabilities.
\end{remark}

\begin{proof}

Let $\widehat{P}$ denote the empirical probability measure from the
$k$ samples.  Azuma's inequality gives that if $Y\sim
\hbox{Bin}(k,p)$ then
\[
P(|Y-kp| > \gamma k) \leq 2\exp(-2\gamma^2 k)
\]
and so for any collection $U=\{u_1,\ldots,u_{l}\} \subseteq V$ and
$x_1,\ldots,x_{l} \in \mathcal{A}$ we have
\begin{equation}\label{e:chernoff}
P\left( \left| \widehat{P} (X(U)=x_U) - P(X(U)=x_U) \right| \leq
\gamma \right) \leq 2\exp(-2\gamma^2 k).
\end{equation}
There are $A^l {n\choose l}\leq A^l n^l$ such choices of
$u_1,\ldots,u_{l}$ and $x_1,\ldots,x_{l}$.  An application of the
union bound implies that with probability at least $1-A^l n^l
2\exp(-2\gamma^2 k)$ it holds that
\begin{equation}\label{e:probabilityBoundNew}
\left| \widehat{P} (X(U)=x_U) - P(X(U)=x_U) \right| \leq \gamma
\end{equation}
for all $\{u_i\}_{i=1}^l$ and $\{x_i\}_{i=1}^l$. If we additionally
have $l\leq d+2$ and $k\geq C(\gamma) d \log n$, then equation
\eqref{e:probabilityBoundNew} holds with probability at least
$1-A^{d+2} n^{d+2} 2/n^{2\gamma^2C(\gamma) d}$. Choosing $C(\gamma)=
\frac{d+2}{\gamma^2 2d}+C_1$, equation \eqref{e:probabilityBoundNew}
holds with probability at least $1-2A^{d+2}/n^{C_1}$.

For the remainder of the proof assume \eqref{e:probabilityBoundNew} holds.
Taking
\begin{equation}\label{e:defineGamma}\gamma(\epsilon,\delta)=\eps \delta^2/9\, ,\end{equation}
we can bound the error in conditional probabilities as
 \begin{align} \nonumber
&|\widehat{P}(X(v)=x_v|X(U)=x_U) -P(X(v)=x_v|X(U)=x_U)|
\\&=\nonumber
    \left|\frac{\widehat{P}(X(v)=x_v,X(U)=x_U)}{\widehat{P}(X(U)=x_U)}
    - \frac{P(X(v)=x_v,X(U)=x_U)}{P(X(U)=x_U)}\right|
\\&\leq \nonumber \left|\frac{\widehat{P}(X(v)=x_v,X(U)=x_U)}{P(X(U)=x_U)}
    - \frac{P(X(v)=x_v,X(U)=x_U)}{P(X(U)=x_U)}\right|
\\ &\quad+ \nonumber
    \left| \frac{1}{\widehat{P}(X(U)=x_U)}- \frac{1}{P(X(U)=x_U)}\right|
\\ &\leq \frac{\gamma}{\delta} + \frac{\gamma}{(\delta - \gamma)\delta}\label{e:conditionalEstimationError1}
\leq \frac{\eps
\delta^2}{9\delta}+\frac{\eps\delta^2}{9(\delta-\frac{\eps\delta^2}{9})\delta}=
\frac{\eps\delta}{9}+ \frac{\eps}{(9-\eps\delta)}<\frac{\eps}{4}\,
.\end{align}

For each vertex
$v\in V$ we consider all candidate neighborhoods for $v$, subsets
$U\subset V-\{v\}$ with $|U|\leq d$.  The estimate \eqref{e:conditionalEstimationError1} and the triangle inequality imply that if
$N(v)\subseteq U$ then by the Markov property,
\begin{align}\label{e:conditionA1New}
&\big|\widehat{P}(X(v)=x_v|X(U)=x_{U},X(w)=x_w)\nonumber\\
&\quad-\widehat{P}(X(v)=x_v|X(U)=x_{U},X(w)=x_w')\big| < \epsilon/2
\end{align}
for all $w\in V$ and $x_1,\ldots,x_l,x_w,x_w',x_v\in \mathcal{A}$
such that
\begin{align}\label{e:conditionB1New}
\left|\widehat{P}(X(U)=x_{U},X(w)=x_w)\right|
> \delta/2,\nonumber\\
\left|\widehat{P}(X(U)=x_{U},X(w)=x_w')\right|
> \delta/2.
\end{align}
Conversely by conditions \eqref{e:conditionANew} and
\eqref{e:conditionBNew} and the estimate \eqref{e:conditionalEstimationError1}, we have that for
any $U$ with $N(v)\nsubseteq U$ there exists some $w\in V$ and
$x_{u_1},\ldots,x_{u_l},x_w,x_w',x_v\in \mathcal{A}$ such that
equation \eqref{e:conditionB1New} holds but equation
\eqref{e:conditionA1New} does not hold.  Thus, choosing the smallest set $U$ such that \eqref{e:conditionA1New} holds gives the correct neighborhood.

To summarize, with number of samples $$k=\left(\frac{81(d+2)}{\eps^2 \delta^4 2d}+C_1\right) d \log n$$ the algorithm correctly determines the graph $G$ with probability $$P(\widehat{G}(X)=G)\geq 1-2A^{d+2}/n^{C_1}\, .$$

The analysis of the running time is straightforward. There are $n$ nodes, and
for each node we consider $O(n^d)$ neighborhoods. For each candidate neighborhood,
we check approximately $O(n)$ nodes and perform a correlation test of complexity
$O(\log n)$.

\end{proof}

\subsection{General  Reconstruction}

While Theorem \ref{t:reconstructionNew} applies to a wide range of
models, condition \eqref{e:conditionANew} may occasionally be too restrictive.
One setting in which condition \eqref{e:conditionANew} does not
apply is if the marginal spin at some vertex $v$ is independent of
the marginal spins at all its neighbors, (i.e for all $u \in N(v)$
and all $x,y\in \mathcal{A}$ we have
$P(X(v)=x,X(u)=y)=P(X(v)=x)P(X(u)=y)$. In this case the algorithm
would incorrectly return the empty set for the neighborhood of $v$.
The weaker conditions for Theorem \ref{t:reconstruction} hold on
essentially all Markov random fields. In particular,
condition~(\ref{e:conditionA}) says that the potentials are non-degenerate, which is clearly a necessary condition in order to recover the graph. Condition~(\ref{e:conditionB}) holds for many models, for example all models with
soft constraints. This additional generality comes at a computational cost, with the algorithm for Theorem 2
having a faster running
time, $O(n^{d+2}\log n)$ versus $O(n^{2d+1}\log n)$.

\begin{theorem}\label{t:reconstruction}
For an assignment $x_U=(x_{u_1},\dots,x_{u_l})$ and $x_{u_i}'\in \mathcal{A}$,
define
\[
x^{i}_U(x_{u_i}') = (x_{u_1}, \dots,x_{u_i}', \dots, x_{u_l})
\]
to be the assignment obtained from $x_U$ by replacing the $i$th element by $x_{u_i}'$.
Suppose there exist $\epsilon,\delta>0$ such that the following
condition holds: for all $v\in V$, if $N(v)=u_1,\ldots,u_l$, then
for each $i,1\leq i\leq l$ and for any set $W\subset V-(v\cup N(v))$
with $|W|\leq d$ there exist values
$x_v,x_{u_1},\ldots,x_{u_i},\dots,x_{u_l},x_{u_i}'\in \mathcal{A}$
and $x_W\in  \mathcal{A}^{|W|}$ such that
\begin{equation}\begin{split}\label{e:conditionA}
&\big|P(X(v)=x_v|X(N(v))=x_{N(v)})
\\&\quad
    -P(X(v)=x_v|X(N(v))=x_{N(v)}^i(x_{u_i}'))\big| >\epsilon
\end{split}\end{equation}
and
\begin{equation}\begin{split}\label{e:conditionB}
&\left|P( X(N(v))=x_{N(v)},X(W)=x_W)\right|
> \delta,
\\
&\left|P(X(N(v))=x_{N(v)}^i,X(W)=x_W)\right|
> \delta.
\end{split}\end{equation}
Then for some constant $C=C(\epsilon,\delta)>0$, if $k>Cd\log n$
then there exists an estimator $\G(\underline{X})$ such that the probability of correct reconstruction is
$P(G=\G(\underline{X}))=1-o(1)$.
The estimator $\G$ is computable in time $O(n^{2d+1} \log n)$.
\end{theorem}


\begin{proof}

As in Theorem \ref{t:reconstructionNew} we can assume that with high
probability we have
\begin{equation}\label{e:probabilityBound}
\left| \widehat{P} (X(U)=x_U) -
P(X(U)=x_U) \right| \leq \gamma
\end{equation}
for all $\{u_i\}_{i=1}^l$ and $\{x_i\}_{i=1}^l$ when  $l\leq 2d+1$
and $k\geq C(\gamma) d \log n$ so we assume that
\eqref{e:probabilityBound} holds. For each vertex $v\in V$ we
consider all candidate neighborhoods for $v$, subsets
$U=\{u_1,\ldots,u_l\}\subset V-\{v\}$ with $0\leq l\leq d$. For each
candidate neighborhood $U$, the algorithm computes a score
\begin{equation}\begin{split}f(v;U)&=\\
\min_{W,i} \max_{x_v,x_W,x_U,x_{u_i}'} &\big|\widehat{P}(X(v)=x_v|X(W)=x_W,X(U)=x_U)\nonumber\\
&-\widehat{P}(X(v)=x_v|X(W)=x_W,X(U)=x_U^i(x_{u_i}'))\big|,\end{split}
\end{equation}
where for each $W,i$, the maximum is taken over all $x_v,X_W,x_U,x_{u_i}'$, such
that
\begin{align}\label{e:lBoundProbability}
\widehat{P}(X(W)=x_W,X(U)=x_U)>\delta/2\\
\widehat{P}(X(W)=x_W,X(U)=x_U^i(x_{u_i}'))>\delta/2\nonumber
\end{align}
and $W\subset V-(\{v\}\cup U)$ is an arbitrary set of nodes of
size $d$, $x_W\in \mathcal{A}^d$ is an arbitrary assignment of
values to the nodes in $W$, and $1\leq
i \leq l$.

The algorithm selects as the neighborhood of $v$ the largest set
$U\subset V-\{v\}$ with $f(v;U)>\eps/2$. It is necessary to check
that if $U$ is the true neighborhood of $v$, then the algorithm
accepts $U$, and otherwise the algorithm rejects $U$.

Taking $\gamma(\epsilon,\delta)=\eps \delta^2/9$, it follows exactly as in Theorem~\ref{t:reconstructionNew} that
the error in each of the relevant empirical conditional probabilities satisfies
 \begin{align} \nonumber
&|\widehat{P}(X(v)=x_v|X(W)=x_W,X(U)=x_U) \\ &\quad
-P(X(v)=x_v|X(W)=x_W,X(U)=x_U)| <\frac{\eps}{4}\,
.\label{e:conditionalEstimationError2}\end{align} If $U\nsubseteq
N(v)$, choosing $u_i\in U- N(v)$, we have when $N(v)\subset W\cup U$
\begin{align*}
&\big|P(X(v)=x_v|X(W)=x_W,X(U)=x_U)
-P(X(v)=x_v|X(W)=x_W,X(U)=x_U^i(x_{u_i}'))\big|
\\&= \big|P(X(v)=x_v|X(N(v))=x_{N(v)})-P(X(v)=x_v|X(N(v))=x_{N(v)})\big|\\ &=0\, ,\end{align*}
by the Markov property \eqref{e:MarkovProperty2}. Assuming that equation \eqref{e:probabilityBound} holds with $\gamma$ chosen as in \eqref{e:defineGamma}, the estimation error in $f(v;U)$ is at most $\eps/2$ by equation \eqref{e:conditionalEstimationError2}, and it holds that $f(v;U)<\eps/2$ for each $U\nsubseteq N(v)$.
Thus all $U\nsubseteq N(v)$ are rejected.
If $U=N(v)$, then by the Markov property \eqref{e:MarkovProperty2} and the conditions \eqref{e:conditionA} and \eqref{e:conditionB}, for any $i$ and $W\subset V$,
\begin{align*}
&\big|P(X(v)=x_v|X(W)=x_W,X(U)=x_U)-P(X(v)=x_v|X(W)=x_W,X(U)=x_U^i(x_{u_i}'))\big|
\\&= \big|P(X(v)=x_v|X(N(v))=x_{N(v)})-P(X(v)=x_v|X(N(v))=x_{N(v)}^i(x_{u_i}'))\big|\\&>\eps\, \end{align*}
for some $x_v,x_W,x_U,x_{u_i}'$.  The error in $f(v;U)$ is less than
$\eps/2$ as before, hence $f(v;U)> \eps/2$ for $U=N(v)$.  Since
$U=N(v)$ is the largest set that is not rejected, the algorithm
correctly determines the neighborhood of $v$ for every $v\in V$ when
\eqref{e:probabilityBound} holds.

To summarize, with number of samples $$k=\left(\frac{81(2d+1)}{\eps^2 \delta^4 2d}+C_1\right) d \log n$$ the algorithm correctly determines the graph $G$ with probability $$P(\widehat{G}(X)=G)\geq 1-2A^{2d+1}/n^{C_1}\, .$$

The analysis of the running time is similar to the previous algorithm.
\end{proof}

\subsection{Non-degeneracy of Models}

We can expect conditions \eqref{e:conditionA} and
\eqref{e:conditionB} to hold in essentially all models of interest.
The following proposition shows that they hold for any model with
soft constraints.

\begin{proposition}[Models with soft constraints]
In a graphical model with maximum degree $d$ given by equation
\eqref{e:gibbsDist} suppose that all the potentials $\Psi_{uv}$
satisfy $\|\Psi_{uv}\|_\infty \leq K$ and
\begin{align}\label{e:softConstraints}
\max_{x_1,x_2,x_3,x_4\in\mathcal{A}} \left| \Psi_{uv}(x_1,x_2) -
\Psi_{uv}(x_3,x_2)-\Psi_{uv}(x_1,x_4) +
\Psi_{uv}(x_3,x_4)\right|>\gamma,
\end{align}
for some $\gamma>0$.  Then there exist $\epsilon,\delta>0$ depending
only on $d,K$ and $\gamma$ such that the hypothesis of Theorem
\ref{t:reconstruction} holds.

\end{proposition}

\begin{proof}
It is clear that for some sufficiently small
$\delta=\delta(d,m,K)>0$ we have that for all
$u_1,\ldots,u_{2d+1}\in V$ and $x_{u_1},\ldots,x_{u_{2d+1}}\in
\mathcal{A}$ that
\begin{equation}\label{e:softLBound}
P(X(u_1)=x_{u_1},\dots,X(u_{2d+1})=x_{u_{2d+1}}) > \delta.
\end{equation}
Now suppose that $u_1,\ldots,u_l$ is the neighborhood of $v$.  Then
for any $1\leq i\leq l$ it follows from equation
\eqref{e:softConstraints} that there exists
$x_v,x_v',x_{u_i},x_{u_i}'\in\mathcal{A}$ such that for any
$x_{u_1}\ldots,x_{u_{i-1}},x_{u_{i+1}},\ldots,x_{u_{l}}\in\mathcal{A}$,
\begin{align*}
\frac{P(X(v)=x_v|X(u_1)=x_{u_1},\dots,X(u_i)=x_{u_i}',\dots,X(u_l)=x_{u_l})}{P(X(v)=x_v'|X(u_1)=x_{u_1},\dots,X(u_i)=x_{u_i}',\dots,X(u_l)=x_{u_l})}\\
\geq
e^\gamma\frac{P(X(v)=x_v|X(u_1)=x_{u_1},\dots,X(u_i)=x_{u_i},\dots,X(u_l)=x_{u_l})}{P(X(v)=x_v'|X(u_1)=x_{u_1},\dots,X(u_i)=x_{u_i},\dots,X(u_l)=x_{u_l})}.
\end{align*}
Combining with equation \eqref{e:softLBound}, condition
\eqref{e:conditionA} follows.
\end{proof}

Although the results to follow hold more generally, for ease of
exposition we will keep in mind the example of the Ising model with
no external magnetic field,
 \begin{equation}\label{e:IsingModel}
  P(\vec{x})=\frac{1}{Z}\exp\left(\sum_{(u,v)\in E}\beta_{uv}x_u x_v\right)\, ,
\end{equation}
where $\beta_{uv}\in \R$ are coupling constants and $Z$ is a normalizing constant.

The following lemma gives explicit bounds on $\eps,\delta$ in terms of bounds on the coupling constants in the Ising model, showing that the conditions of Theorem \ref{t:reconstruction} can be expected to hold quite generally.

\begin{proposition}\label{l:IsingSatisfiesConditions}
  Consider the Ising model with all parameters satisfying $$0<c<|\beta_{ij}|<C$$ on a graph $G$ with max degree at most $d$ . Then the conditions \eqref{e:conditionA} and \eqref{e:conditionB} of Theorem~\ref{t:reconstruction} are satisfied with $$\eps \geq \frac{\tanh(2c)}{2C^2+2C^{-2}}$$ and $$\delta \geq \frac{e^{-4dC}}{2^{2d}} .$$
\end{proposition}
\begin{proof}
  Fix a vertex $v\in V$ and let $w\in N(v)$ be any vertex in the neighborhood of $v$. Let $R= N(v)\setminus \{w\}$ be the other neighbors of $v$. Then
  \begin{equation}\begin{split}\label{e:conditionalProbabilityExpansion}
    &P(X(v)=1|X(R)=x_R,X(w)=x_w)\\&= \frac{P(X(v)=1,X(R)=x_R,X(w)=x_w)}{P(X(v)=1,X(R)=x_R,X(w)=x_w)+P(X(v)=0,X(R)=x_R,X(w)=x_w)}
    \\&= \frac{\exp\left(\sum_{j\in R} x_j \beta_{jv}+x_w \beta_{wv}\right)}{\exp\left(\sum_{j\in R} x_j \beta_{jv}+x_w \beta_{wv}\right)+\exp\left(-\sum_{j\in R} x_j \beta_{jv}-x_w \beta_{wv}\right)}.
  \end{split}\end{equation}
  Defining $$A:= \exp\left(\sum_{j\in R} x_j \beta_{jv}\right),$$  we have from \eqref{e:conditionalProbabilityExpansion} that
  \begin{align*}
    &|P(X(v)=1|X(R)=x_R,X(w)=1)-P(X(v)=1|X(R)=x_R,X(w)=-1)|
    \\ &\quad =\left| \frac{Ae^{\beta_{wv}}}{Ae^{\beta_{wv}}+ A^{-1} e^{-\beta_{wv}}} - \frac{Ae^{-\beta_{wv}}}{Ae^{-\beta_{wv}}+A^{-1}e^{\beta_{wv}}}\right|
    \\ &\quad = \left| \frac{A^2 (e^{2\beta_{wv}}-e^{-2\beta_{wv}})}{A^4+A^2(e^{2\beta_{wv}}+e^{-2\beta_{wv}})+1}\right|
    \\ &\quad = \frac{A^2 (e^{2|\beta_{wv}|}-e^{-2|\beta_{wv}|})}{A^4+A^2(e^{2|\beta_{wv}|}+e^{-2|\beta_{wv}|})+1}
    \\ &\quad = \frac{(e^{2|\beta_{wv}|}-e^{-2|\beta_{wv}|})}{A^2+e^{2|\beta_{wv}|}+e^{-2|\beta_{wv}|}+A^{-2}}
  \geq \frac{\tanh(2|\beta_{wv}|)}{2A^2+2A^{-2}}\, .
    \end{align*}
    It is possible to choose the spins $x_R$ in such a way that $e^{-C}<A<e^{C}$. Thus the expression above is at least
    $$\frac{\tanh(2c)}{2e^{2C}+2e^{-2C}}
    \, .$$
    Moreover, the probability of any assignment of $2d$ spins can be very crudely bounded as $$P(X(i_1)=x_{i_1},\dots,X(i_{2d})=x_{i_{2d}})\geq \frac{e^{-4dC}}{2^{2d}}\, .$$

\end{proof}

\subsection{$O(n^2 \log n)$  Algorithm For Models with Correlation Decay}

The reconstruction algorithm runs in polynomial time $O(d
n^{2d+1}\ln n)$. It would be desirable for the degree of the
polynomial to be independent of $d$ and this can be achieved for
Markov random fields with exponential decay of correlations.  For
two vertices $u,v\in V$ let $d(u,v)$ denote the graph distance and
let $d_C(u,v)$ denote the correlation between the spins at $u$ and
$v$ defined as
\[
d_C(u,v) = \sum_{x_u,x_v\in\mathcal{A}} \left|
P(X(u)=x_u,X(v)=x_v)-P(X(u)=x_u)P(X(v)=x_v)\right|.
\]
If the interactions are sufficiently weak the graph will satisfy the
Dobrushin-Shlosman condition (see e.g. \cite{Dobrushin}) and there
will be exponential decay of correlations between vertices.

\begin{theorem}\label{t:reconstructionDecay}
Suppose that $G$ and $X$ satisfy the hypothesis of Theorem
\ref{t:reconstruction} and that for all $u,v\in V$, $d_C(u,v)\leq
\exp(-\alpha d(u,v))$ and there exists some $\kappa>0$ such that for
all $(u,v)\in E$, $d_C(u,v)>\kappa$. Then for some constant
$C=C(\alpha,\kappa,\epsilon,\delta)>0$, if $k>Cd\log n$ then there
exists an estimator $\G(\underline{X})$ such that the probability of
correct reconstruction is $P(G=\G(\underline{X}))=1-o(1)$ and the
algorithm runtime is $O(nd^{\frac{d\ln(4/\kappa)}{\alpha}}+d n^2\ln
n)$ with high probability.
\end{theorem}

\begin{proof}
Denote the correlation neighborhood of a vertex $v$ as
$N_C(v)=\{u\in V: \widehat{d_C}(u,v)>\kappa/2\}$ where
$\widehat{d_C}(u,v)$ is the empirical correlation of $u$ and $v$.
For large enough $C$ with high probability for all $v\in V$ we have
that $N(v)\subseteq N_C(v) \subseteq \{u\in V:d(u,v)\leq
\frac{\ln(4/\kappa)}{\alpha} \}$.  Now the size of $|\{u\in
V:d(u,v)\leq \frac{\ln(4/\kappa)}{\alpha} \}|\leq
d^{\frac{\ln(4/\kappa)}{\alpha}}$ which is independent of $n$.

When reconstructing the neighborhood of a vertex $v$ we modify the
algorithm in Theorem \ref{t:reconstruction} to only test candidate
neighborhoods $U$ and sets $W$ which are subsets of $N_C(v)$.  The
algorithm restricted to the smaller range of possible neighborhoods
correctly reconstructs the graph since the true neighborhood of a
vertex is always in its correlation neighborhood.  For each  vertex
$v$ the total number of choices of candidate neighborhoods $U$ and
sets $W$ the algorithm has to check is
$O(d^{\frac{d\ln(4/\kappa)}{\alpha}})$ so running the reconstruction
algorithm takes $O(n d^{\frac{d\ln(4/\kappa)}{\alpha}})$ operations.
It takes $O(d n^2\ln n)$ operations to calculate all the
correlations which for large $n$ dominates the run time.
\end{proof}

\section{Noisy and Incomplete Observations} \label{sec:noisyetc}

More generally there is the problem of reconstructing a Markov
random field from noisy observations.  In this setting we observe
$\underline{Y}=\{Y^1,\ldots,Y^k\}$ instead of
$\underline{X}=\{X^1,\ldots,X^k\}$ where each $Y_i$ is a noisy
version of $X_i$.  The algorithm in Theorem \ref{t:reconstruction}
is robust to small amounts of noise, even when the errors in
different vertices are not necessarily independent. One sufficient
condition is that there exist $0<\epsilon'<\epsilon$ and
$0<\delta'<\delta$ such that for any $2d+1$ vertices
$v_1,\ldots,v_{2d+1}$ and states $x_1,\ldots,x_{2d+1}$ we have that
\[
\left| P(X(v_1)=x_1,\ldots,X(v_{2d})=x_{2d})-
P(Y(v_1)=x_1,\ldots,Y(v_{2d})=x_{2d}) \right|\leq \delta'/2
\]
and
\begin{equation*}\begin{split}
&\big|P(X(v_{2d+1})=x_{2d+1}|X(v_1)=x_1,\ldots,X(v_{2d})=x_{2d})
\\&\quad
    -P(Y(v_{2d+1})=x_{2d+1}|Y(v_1)=x_1,\ldots,Y(v_{2d})=x_{2d})\big| \leq
    \epsilon'/2.
\end{split}\end{equation*}
For some $C'=C'(\epsilon,\epsilon',\delta,\delta')>0$ with
$k=C'd\log n$ samples the reconstruction algorithm of Theorem \ref{t:reconstruction} correctly
reconstructs the graph $G$ with high probability (the same
proof holds).

\subsection{An Example of Non-Identifiability} \label{subsec:nonid}

Without assumptions on the underlying model or noise, the
Markov random field is not in general identifiable. In other words, a single probability distribution might correspond to two different graph structures. Thus, the problem of reconstruction is not well-defined in such a case. The next example shows that even in the
Ising model, under unknown noise it is impossible to distinguish between a
graph with 3 vertices and 2 edges and a graph with 3 vertices and 3
edges.

\begin{example}

Let $V=\{v_1,v_2,v_3\}$ be a set of 3 vertices and let $G$ and
$\widetilde{G}$ be two graphs with vertex set $V$ and edge sets
$\{(u_1,u_2),(u_1,u_3)\}$ and $\{(u_1,u_2),(u_1,u_3),(u_2,u_3)\}$
respectively.  Let $P$ and $\widetilde{P}$ be Ising models on $G$
and $\widetilde{G}$ with edge interactions $\beta_{12},\beta_{13}$
and
$\widetilde{\beta_{12}},\widetilde{\beta_{13}},\widetilde{\beta_{23}}$
respectively, i.e.
\begin{align*}P[X]&= \frac1{Z}\exp\left(\beta_{12}X(u_1)X(u_2)+\beta_{13}
X(u_1)X(u_3)\right) \\
\widetilde{P}[X]&= \frac1{Z}\exp\left(\tbeta_{12}X(u_1)X(u_2)+\tbeta_{13}
X(u_1)X(u_3)+\tbeta_{23} X(u_2) X(u_3)\right).
\end{align*}
Suppose that $X'(u_1)$, a noisy version of the spin $X(u_1)$, is
observed which is equal to $X(u_1)$ with probability $p$ and
$-X(u_1)$ with probability $1-p$ for some random unknown $p$ while
the spins $X(u_2)$ and $X(u_3)$ are observed perfectly.  This is
equivalent to adding a new vertex $u_1'$ to $G$ and $\widetilde{G}$
with an extra edge $(u_1,u_1')$ and potential
$\Psi_{(u_1,u_1')}=\beta_{11'}X(u_1)X(u_1')$.  The spin at $u_1'$
then represents the noisy observation of the spin at $u_1$.  Suppose
that all the $\beta$ and $\widetilde{\beta}$ are chosen
independently with $N(0,1)$ distribution and let $\mathcal{P}$ and
$\widetilde{\mathcal{P}}$ be the random noisy distributions on
$\mathcal{A}^{\{u_1',u_2,u_3\}}$. Then the total variation distance
between $\mathcal{P}$ and $\widetilde{\mathcal{P}}$ is less than 1
and so the graph structure is not identifiable as we shall show
below.

By the symmetry of the Ising model with no external field the random
element $\mathcal{P}$ can be parameterized by
$(p_{1'2},p_{1'3},p_{23})\in [0,1]^3$ where
$p_{1'2}=P(X_{u_1'}=1,X_{u_2}=1),p_{1'3}=P(X_{u_1'}=1,X_{u_3}=1),
p_{23}=P(X_{u_2}=1,X_{u_3}=1)$.  These parameters are given by
\[
p_{ij}=h(\beta_{1i})h(\beta_{1j})+h(-\beta_{1i})h(-\beta_{1j})
\]
where $h(\beta)=\frac{e^\beta}{e^\beta+e^{-\beta}}$.  Let $\phi$ be
the function $\phi:\mathbb{R}^3\rightarrow [0,1]^3$ which maps
$(\beta_{11'},\beta_{12},\beta_{13}) \mapsto
(p_{1'2},p_{1'3},p_{23})$ and let $J_\phi$ be its Jacobian.  Then
$\det (J_\phi(1,1,1))>0$ and by continuity the Jacobian is positive
in a neighborhood of $(1,1,1)$.  It follows that the random vector
$(p_{1'2},p_{1'3},p_{23})$ has a density with respect to Lebesgue
measure in a neighborhood of $(2h(1)^2,2h(1)^2,2h(1)^2)$.

Now let $\widetilde{\phi}$ be the function
$\widetilde{\phi}:\mathbb{R}^3\rightarrow [0,1]^3$ which maps
$(\widetilde{\beta_{11'}},\widetilde{\beta_{12}},\widetilde{\beta_{13}},\widetilde{\beta_{23}})\mapsto
(\widetilde{p_{1'2}},\widetilde{p_{1'3}},\widetilde{p_{23}})$.  If
we fix $\widetilde{\beta_{23}}=0$ then $\widetilde{\phi}=\phi$
induces a positive density in the random vector
$(\widetilde{p_{1'2}},\widetilde{p_{1'3}},\widetilde{p_{23}})$ in a
neighborhood of $(2h(1)^2,2h(1)^2,2h(1)^2)$.  By continuity this
also holds when $|\widetilde{\beta_{23}}|$ is small enough  and so
$(\widetilde{p_{1'2}},\widetilde{p_{1'3}},\widetilde{p_{23}})$ has a
positive density around $(2h(1)^2,2h(1)^2,2h(1)^2)$.  Hence we have
that both $\mathcal{P}$ and $\widetilde{\mathcal{P}}$ have positive
densities in an overlapping region so their total variation distance
is less than 1 and so the graph structure is not identifiable.
\end{example}

\subsection{Models With Hidden Variables}
A related question is can we identify if a vertex is missing and if
so where it fits into the graph.  Under the assumption that the
vertices all have degree at least 3 and the graph is triangle-free
we can recover missing vertices under mild assumptions.

\begin{theorem}\label{t:missing_vertex}
Suppose that the hypothesis of Theorem \ref{t:reconstruction} holds
for some Markov random field $X$ based on a triangle-free  graph
with minimum degree at least 3 and maximum degree $d'$. Let
$V^*\subseteq V$ such that for any two points $v,v'\in V-V^*$ we
have $d(v,v')\geq 3$ and suppose we are given samples from $X^*$,
the restriction of $X$ to $V^*$ with which to reconstruct $G$.

Suppose the following condition also holds:  for all $v\in V$ if
$v_1,v_2\in N(v)$ and $U=N(v)\cup N(v_1)-\{v,v_1,v_2\}$ and
$W\subset V-(N(v)\cup N(v_1))$ with $|W|\leq 2d$ then there exists
some $x_{v_1},x_{v_2},x_{v_2}',x_U,x_W$ such that
\begin{equation}\begin{split}\label{e:conditionAmissing}
&\big|P(X(v_1)=x_{v_1}|X(W)=x_{W},X(U)=x_{U},X(v_2)=x_{v_2})
\\&\quad
    -P(X(v_1)=x_{v_1}|X(W)=x_{W},X(U)=x_{U},X(v_2)=x_{v_2}')\big| >\epsilon
\end{split}\end{equation}
and
\begin{equation}\begin{split}\label{e:conditionBmissing}
&\left|P(X(W)=x_{W},X(U)=x_{U},X(v_2)=x_{v_2})\right|
> \delta,
\\
&\left|P(X(W)=x_{W},X(U)=x_{U},X(v_2)=x_{v_2}')\right|
> \delta.
\end{split}\end{equation}
Then for some constant $C=C(\epsilon,\delta)>0$, if $k>Cd\log n$
then there exists an estimator $\G(\underline{X}^*)$ such that the
probability of correct reconstruction is
$P(G=\G(\underline{X}^*))=1-o(1)$.
\end{theorem}

\begin{proof}
We apply the algorithm from Theorem \ref{t:reconstruction} to
$\underline{X}^*$ setting the maximum degree as $d=2d'$.  The
algorithm will output the graph $G^*=(V^*,E^*)$.  If $v,N(v)\subset
V^*$ then the algorithm correctly reconstructs the neighborhood
$N(v)$.  Any vertex in $V^*$ is adjacent to at most one missing
vertex so suppose that $v_1$ is a vertex adjacent to a missing
vertex $v$.  Then by condition \eqref{e:conditionAmissing} and
\eqref{e:conditionBmissing} we have that the algorithm reconstructs
the neighborhood of $v_1$ as $N(v)\cup N(v_1)-\{v,v_1\}$.  So the
edge set $E^*$ is exactly all the edges in the induced subgraph of
$V^*$ plus a clique connecting all the neighbors of missing
vertices.  Since $G$ is triangle-free every maximal clique (a clique
that cannot be enlarged) of size at least 3 corresponds to a missing
vertex.

So to reconstruct $G$ from $G^*$ we simply replace every maximal
clique in $G^*$ with a vertex connected to all the vertices in the
clique.  This exactly reconstructs the graph with high probability.

\end{proof}

\begin{remark}The condition that missing vertices are at distance at least 3 is
not necessary, but this assumption simplifies the algorithm because the cliques
corresponding to missing vertices are disjoint.  A slightly more involved
algorithm is able to reconstruct graphs where the missing vertices have $d(v,v')=2$.
\end{remark}

The following lemma shows that the conditions for recovery of missing vertices in Theorem
\ref{t:missing_vertex} are satisfied for a ferromagnetic Ising model
satisfying the assumptions of Lemma~\ref{l:IsingSatisfiesConditions}.
\begin{lemma}\label{l:missing_vertex_monotonic}
Consider the ferromagnetic Ising model where all coupling parameters satisfy $$0<c<\beta_{ij}<C\,$$ on a triangle-free graph $G$ with minimum degree 3. Then the conditions of Theorem~\ref{t:missing_vertex} are satisfied with $$\eps\geq \frac{\tanh(2c)}{32 e^{2(d+1)C}(C^2+C^{-2})} ,$$ and $$\delta\geq \frac{e^{-4dC}}{2^{2d}} .$$
\end{lemma}
\begin{proof}To check the first condition we write
\begin{align*}
    &\big|P(X(v_1)=1|X(N)=x_N,X(v_2)=1)
    \\ &\quad - P(X(v_1)=1|X(N)=x_N,X(v_2)=-1)\big|
    \\ & = \big|P(X(v_1)=1|X(N)=x_N,X(v_2)=1,v=1)P(v=1|X(N)=x_N,X(v_2)=1)
    \\& \quad + P(X(v_1)=1|X(N)=x_N,X(v_2)=1,v=-1)P(v=-1|X(N)=x_N,X(v_2)=1)
    \\ &\quad - P(X(v_1)=1|X(N)=x_N,X(v_2)=-1,v=1)P(v=1|X(N)=x_N,X(v_2)=-1)
    \\ &\quad - P(X(v_1)=1|X(N)=x_N,X(v_2)=-1,v=-1)P(v=-1|X(N)=x_N,X(v_2)=-1)\big|
    \\&=
        \big|P(X(v_1)=1|X(N)=x_N,v=1)P(v=1|X(N)=x_N,X(v_2)=1)
    \\& \quad + P(X(v_1)=1|X(N)=x_N,v=-1)P(v=-1|X(N)=x_N,X(v_2)=1)
    \\ &\quad - P(X(v_1)=1|X(N)=x_N,v=1)P(v=1|X(N)=x_N,X(v_2)=-1)
    \\ &\quad - P(X(v_1)=1|X(N)=x_N,v=-1)P(v=-1|X(N)=x_N,X(v_2)=-1)\big|\,
  \end{align*}
where $N=N(v)\cup N(v_1)-\{v,v_1,v_2\}$ and where the last step
follows by the Markov property (since all paths from $v_1$ to $v_2$
pass through vertices in $N$ or through $v$). Continuing, we have
that the above is equal to
  \begin{equation}
    \left| \left(P(v_1=1|N,v=1)-P(v_1=1|N,v=-1)\right)\left(P(v=1|N,v_2=1)-P(v=1|N,v_2=-1)\right)\right|\, .
  \end{equation}
But by Lemma~\ref{l:IsingSatisfiesConditions}, $$\big|
\left(P(v_1=1|N,v=1)-P(v=1|N,v=-1)\right)\big|>\frac{\tanh(2c)}{2C^2+2C^{-2}}\,
.$$ By the ferromagnetic assumption, the second factor can be lower
bounded as $$\big|\left(P(v=1|N,v_2=1)-P(v=1|N,v_2=-1)\right)\big|>
\frac{1}{16 e^{2(d+1)C}} \,.$$ Hence the first condition is
satisfied with $$\eps>\frac{\tanh(2c)}{32
e^{2(d+1)C}(C^2+C^{-2})}\,.$$
The second condition, by the same
argument as Lemma~\ref{l:IsingSatisfiesConditions}, is satisfied
with $$\delta\geq \frac{e^{-4dC}}{2^{2d}}\,.$$
\end{proof}

\paragraph{Acknowledgment}
E.M. thanks Marek Biskup for helpful discussions on models with hidden variables.

\end{document}